# Nature evolution of the shortened bond in atomic clusters and at junction interfaces

Chang Q. Sun,[1*] C. M. Li,[2] Y. Shi,[3] Z.Q. Li,[4] H. L. Bai[4] and E. Y. Jiang[4]

[1] *School of Electrical and Electronic Engineering, Nanyang Technological University, Singapore 639798*

[2] *School of Chemical and biomedical Engineering, Nanyang Technological University, Singapore 639798*

[3] *East Campus, Tangshan University, Hebei Province, P. R. China 063020*

[4] *Institute of Advanced Materials Physics and Faculty of Science, Tianjin University, P. R. China 300072*

Abstract

Thermally stimulated process such as evaporation, phase transition, or solid-liquid transition of a solid consumes each a certain portion of the solid cohesive energy that is the sum of bond energy over all the coordinates of all the involved atoms. Generally, the critical temperatures for such processes drop with solid size, unless hetero capping or interfacial interaction becomes dominant, because of the increased portion of the lower-coordinated surface atoms [Sun et al., *J. Phys. Chem.* B108, 1080 (2004)]. It is intriguing, however, that the melting point ($T_m$) of a solid containing III-A or IV-A atoms oscillates with size (the $T_m$ drops first and then rises as the solid size is reduced) and that the $T_m$ of chemically capped nanosolid often increases with the inverse size. Here we show that bond nature evolution is essential for the selective nanosolids and at the junction interfaces, which is responsible for the superheating of the smallest nanosolids, chemically capped clusters, and junction interfaces that exhibit insulating nature with high mechanical strength.

# I  Introduction

Thermal stability of a nanosolid indicated by phase transition, coalescence, solid-liquid transition, or evaporation has been a long historical mystery, which has attracted tremendously renascent efforts of investigation towards the mechanism behind.[1,2,3,4,5,6,7,8] It has been well established both experimentally[9,10] and theoretically[11,12] that the critical temperatures ($T_C(K_j)$) for such phase transitions of a freestanding solid in the size range of nanometers drops with size in a $K_j^{-1}$ fashion unless the nanosolid is chemically capped or strongly bonded to the substrate, where $K_j = R_j/d$ is the dimensionless form of size that equals the number of atoms lined along the radius of a spherical dot with radius $R_j$ or cross a thin film of $R_j$ thick and d is the atomic diameter in the bulk. The phenomena of melting point ($T_m(K_j)$) suppression and elevation are often referred as supercooling and superheating, respectively. It has been clear that the melting of impurity-free solid proceeds predominantly in a form of liquid shell nucleation and growth if little defect is involved as it has been confirmed that a flat or curved surface melts easier than the bulk interior.[13] For instance, the surface layer of $Al^+_{51}$ and $Al^+_{52}$ clusters melts 100 K before the cores of the clusters.[14] However, it is surprising that a chemically capped nansolid melts at temperature higher than the bulk value of the core material and that a free-standing nanosolid at the lower end of the size limit, or clusters containing 10 ~ 60 atoms of III-A or IV-A elements, melt at temperatures that are 10 ~ 100% or even higher than the corresponding bulk $T_m(\infty)$.[15,16,17,18,19] For example, $Ga^+_{39-40}$ clusters were measured to melt at about 550 K, while $Ga^+_{17}$ cluster does not melt even up to 700 K compared with the $T_m(\infty)$ of 303 K.[15] Small Sn clusters with 10 ~ 30 atoms melt at least 50 K above the $T_m(\infty)$ of 505 K.[16] Advanced *ab initio* density functional molecular dynamics simulations suggest that $Ga^+_{13}$ and $Ga^+_{17}$ clusters melt at 1400 and 650 K.[17] Calculations even suggested that $Sn_n$ (n = 6, 7, 10, and 13) clusters melt at 1300, 2100, 2000, and 1900 K, respectively.[19] Theoretical calculations also predicted that the structural transition happens at 500 and 1500 K for a $Sn_{10}$ cluster, and for a $Sn_{20}$ cluster it happens at 500 and 1200 K.[20] Recent calorimetric measurements[21] on unsupported $Sn^+_n$ particles clarified that $Sn^+_{10}$ and $Sn^+_{11}$ clusters can sustain till 1073 K while Sn clusters containing n >20 and n < 8 atoms are less thermally stable as the melt occurs at temperature around 773 K or below. $Sn^+_{19}$ can sustain till 673 K while $Sn^+_{20}$ melts below 673 K. Measurements also show that unsupported $Al^+_{49-63}$ clusters, melt in the temperature range between 450 and 650 K (well below the bulk melting point of 933 K). The peaks for $Al^+_{51-52}$ clusters are bimodal, suggesting the presence of a premelting transition where the surface of the clusters melts around 100 K before the core. For clusters with $n > 55$ the melting temperatures suddenly drop, and there is a dip in the heat capacities due to a transition between two solid forms before the clusters melt.[14] Calculations[18] also suggested that the IV-A elements, $C_n$, $Si_n$, $Ge_n$, $Sn_n$ (magic number of n ~13) clusters melt at temperature higher than their bulk $T_m$. The $C_{13}$ cluster prefers a monocyclic ring or a tadpole structure with the most probability to appear in the simulated annealing when the temperature is between 3000 and 3500 K. Although the $T_m$ may be overestimated to some extent for the smallest clusters in simulations,[19] the $T_m$ elevation for smallest clusters happens indeed according to measurement.

The $T_m$ rise of the chemically capped nanosolid is often attributed to the enhancement of interfacial energy.[22,23,24] The $T_m$ elevation of the smallest Ga and Sn nanosolid is attributed either to the bond nature alteration from covalent-metallic to pure covalent with slight bond contraction,[17,25] or to the heavily geometrical reconstruction as Ge, Si, and Sn clusters are found to be stacks of stable tricapped triagonal prism units.[26] However, consistent insight into the $T_m$ oscillation in the whole range of sizes (from single atom to the bulk) and the superheating of the capped nanosolids is yet lacking though numerous outstanding models have been developed specifically for the $T_m$ elevation or suppression. Here we show that the intriguing phenomena of $T_m$ oscillation, suppression, and elevation are within the scope of the recent bond-order-length-strength (BOLS) correlation mechanism,[27] indicating that the competition between the bond order loss and the associated bond strength gain dominates the thermal stability. It has been asserted therefore that bond nature evolution is necessary to the shortened bond in the smallest atomic clusters and at the homo- and hetero-junction interfaces where superheating may happen, forming new insight into the thermal stability, mechanical strength, and electronic properties of the interfaces.

## II  Principle

### 2.1 Extended BOLS correlation

The BOLS correlation mechanism, as detailed in refs 27 and 28, suggests that atomic coordination number (CN, or $z_i$) imperfection of an atom denoted i at site surrounding a defect (voids, stacks of fault, impurities, etc.) or near the surface edge causes the remaining bonds of the lower-coordinated atom to contract (with coefficient $c_i = d_i/d$) from bulk value d to the specific $d_i$ spontaneously associated with bond strength gain. Consequently, the intraatomic potential well is depressed by $E_i/E_b = c_i^{-m}$. The atomic cohesive energy, or the sum of bond energy over all the coordinates, of the lower-coordinated atom will change by $E_{c,i}/E_{c,b} = z_i/z_b c_i^{-m}$. The $E_i$ and $E_b$ are the cohesive energy per bond for the specific ith atom and for an atom in the bulk, respectively. The index m is a key parameter that represents the nature of the bond. It has been verified[27] that for pure metals such as Au, Ag, and Ni, m ≡ 1; for alloys and compounds m is around 4; for C and Si the m has been optimized to be 2.56[29] and 4.88,[30] respectively. Variation of the m value may happen subjecting to bond nature change. The BOLS correlation is expressed as:[27]

$$\begin{cases} c_i(z_i) &= d_i/d = 2/\{1+\exp[(12-z_i)/(8z_i)]\} & (BOLS-coefficient) \\ E_i &= c_i^{-m} E_b & (Single-bond-energy) \\ T_C \propto E_{B,i} &= z_i E_i & (atomic-cohesive-energy) \end{cases}$$

(1)

For a nanosolid, the subscript i denotes an atom in the i th atomic layer, which is countered up to three from the outermost atomic layer to the center of the solid as no CN-reduction is expected for i > 3. Eq (**1**) formulates well, as shown in Figure 1, the bond order-length premises of Goldschmidt[31] and Feibelman,[32] who indicated that the atomic radius will shrink by 3%, 4%, 12%, and 30% when the $z_i$ is reduced from the standard value of 12 to 8, 6, 4, and 2, respectively.

The CN-imperfection induced bond contraction is independent of the nature of the specific bond or structural phases,[33] which happens even at liquid surfaces[34] and at sites surrounding atomic voids[35,36,37] or substitution impurities.[38] Recently, it has been uncovered that the spacing between the first and second atomic surface layers contracts by 10% relative to that of subsequent layers of liquid Sn, Hg, Ga, and Indium.[34] Substitutional doping of Bi and As could induce a 8% contraction of bonds around the As and Bi impurities at the Te sublattices in a CdTe compound.[38] The finding of impurity-induced bond contraction could provide important impact to an atomic scale understanding of the bond in a junction interface that has been puzzled for decades. The discovery of bond contraction at liquid surface could provide a pertinent mechanism for a liquid drop formation and sustention. Therefore, the BOLS correlation can be extended to liquid and junction interfaces such as twin grain boundaries where atomic CN has little change.

As consequences of the BOLS correlation, the CN imperfection induced bond contraction and potential-well depression localize the electrons, which enhances the charge density in the relaxed region. Bond strength gain also enhances the energy density *per unit volume* in the relaxed region, which perturbs the Hamiltonian of an extended solid and the associated properties such as the band and band-gap widths, core-level shift, Stokes shift (electron-phonon interaction), and dielectric suppression. On the other hand, the competition between bond-order loss and the associated bond-strength gain dictates the mechanical strength[35] and thermodynamic process of the solid such as atomic vibration, chemical reactivity, and thermal stability. The CN and m value dependence of the atomic cohesive energy as shown in Figure 1 indicates the possibility of $E_C$ increase at larger m values with CN reduction. The atomic cohesive energy is important as it determines the thermal dynamic behavior of a solid such as phase stability, self-assembly growth, structural deviation, and activation energy for atomic diffusion and dislocation in a solid.

Most strikingly, incorporating the BOLS correlation to the new freedom of physical size has enabled us to elucidate quantitative information of single energy levels of an *isolated* (Si, Pd, Au, Ag, and Cu) atom and its shift upon bulk and nanosolid formation,[27] the vibration frequency of a Si-Si dimer bond,[39] and bonding identities such as the length, strength, extensibility, and thermal and chemical stability, and specific heat of a single bond in the gold monatomic chains[40] and in the carbon nanotubes.[29]

2.2   Correlation between critical temperatures and atomic cohesive energy

Warming up an atom to a critical temperature, or loosening all the bonds of the specific atom to some extent for a specific event, requires energy that is a portion of the atomic cohesive energy,[24,41] $T_C \sim E_C(z_i) = z_i E_b(z_i)$. For a spherical dot with $R_j = K_j d$ radius, the relative changes of the critical temperatures for phase transitions can be unified as:

$$\begin{cases} \dfrac{\Delta T_m(K_j)}{T_m(\infty)} = \dfrac{\Delta T_C(K_j)}{T_C(\infty)} = \dfrac{\Delta T_{vap}(K_j)}{T_{vap}(\infty)} = \dfrac{\Delta E_C(K_j)}{E_C(\infty)} & = \begin{cases} = \sum_3 \gamma_{ij}\left(z_i / z_b c_i^{-m} - 1\right) & (BOLS-prediction) \\ -B/K_j & (Scaling-relation) \end{cases} \\ \gamma_{ij} = N_i/N_j = V_i/V_j = \begin{cases} 1, & K_j < \tau c_i \\ \tau c_i/K_j, & else \end{cases} & (Surface-to-volume-ratio) \end{cases}$$

(2)

where $T_{vap}(K_j)$ is the critical temperature for evaporation. The term $-B/K_j$ is the general scaling relation fitting the measured data of $T_C$ suppression. $\gamma_{ij}$ is the portion of atomic number or volume in the ith atomic layer which is counted from the outmost shell to the center of the solid. The $\tau$ represents the dimensionality of a spherical dot ($\tau = 3$), a rod ($\tau = 2$), and a thin slab ($\tau = 1$). At the lower end of the size limit ($K_j \leq \tau c_i$), $\gamma_{ij} = 1$, which means that all the atoms suffer from CN imperfection. According to the BOLS correlation, the critical temperature for phase transition of an isolated atom is 0 K because $z_i = 0$ (see eq (2)). Experimental explorations[42,43] revealed that the effective CN of a surface atom follows the relation: $z_1 = 4(1-0.75/K)$, and the CN of the subsequent layers are $z_2 = 6$, and $z_3 = 12$.[39] Eq (2) applies not to systems in which long order interaction dominates such as the case of dipole-dipole interaction in ferroelectric systems where the correlation length has to be considered.[28] Equation (2) indicates that the quantity $\alpha = z_i / z_b c_i^{-m}$ which involves the atomic CN, bond length, and the nature of the bond, dictates the process of superheating ($\alpha > 1$, $T_m$ elevation for III-A and IV-A and chemically capped nanosolids) or supercooling ($\alpha < 1$, $T_m$ suppression of a freestanding nanosolid). For a capped nanosolid, $z_i/z_b \sim 1$, the $\alpha$ represents the interfacial bond strengthening as no apparent loss of bond order happens. The only way of $\alpha > 1$ is the m value increases for the shortened interfacial bonds. For a freestanding nanosolid, there are two possibilities for $\alpha > 1$. One is that the m increases as $z_i$ is reduced and the other is that the $c_i$ is much lower than the prediction as given in eq (1). No other sources contribute to the $\alpha$ value according to eq (2).

### III  Results and discussion

3.1  $T_m$ suppression of freestanding nanosolids

Figure 2 compares predictions with the measured $T_m$ suppression of metals, semiconductors, solids of inert gases and methyl-chloride polymer (m-Cl). The measured data are normalized with the corresponding bulk values according to the scaling relations given in eq (2). Acceptable match is gained for all the samples despite the m values and geometrical shapes of the nanosolids. It is interesting to note that Al -01 nanosolids grown on SiN substrate are more plate-like ($\tau = 1$, m = 1) throughout the sizes of measurement (panel a) but Sn-01 on SiN (panel a) and In-01 and Pd-01 (in panel b) are more spherical-like ($\tau = 3$) at particle size smaller than 10 nm. However, the Indium particle (panel b) encapsulated in the controlled-pore silica exhibits slight superheating while the Indium embedded in Vycor glass shows no superheating effect. The melting profiles show that at the smaller size range, the Au/W and Au/C interface promotes more significantly the melting of Au (supercooling) than the silica capped Au that shows slightly superheating (panel c). CdS and Bi nanosolids (panel e) exhibit different modes of melting because of the nanosolid-substrate interaction. The measured $T_m$ of Si-01 (panel d) and CdS (panel e) nanosolids appeared to be lower than the expected trends with m value of 4.88 for Si. The definition of melting is different from source to source, which might be the reason of the deviation between predictions and measurement. For instance, molecular-dynamics calculations revealed that [44] coalescence occurs at temperatures lower than the cluster melting point, and that the difference between coalescence and melting temperatures increases with decreasing cluster size. In the normalization of the scaling relation, the coalescence temperature is lower than the $T_m$ and the coalescence T drops faster than $T_m$ with solid size. The size dependent $T_m$ of Kr, Ne, and O solids follow the curve of m = 4.88 as well, despite the accuracy of measurement. Understanding gained herewith provides not only information about the mode of epitaxial growth but also the bonding status between the nanosolid and the substrate. The deviation of experiment from theory may provide information about the difference in interfacial energy between the particles and the substrates, which is expected to be subject to the temperature of formation.

### 3.2 $T_C$ and $T_{vap}$ suppression

Figure 3 shows the magnetic $T_C$ suppression for Ni thin films deposited on Re and W substrates and the $T_{vap}$ suppression of Ag and PbS nanoparticles. Agreement between measurements and predictions indicates that the relative changes of the critical temperatures are identical regardless of the actual process of melting, evaporating or phase transition that consumes each a certain portion of the solid cohesive energy. The exact portion comes not into play as we seek for the relative change, which forms an additional advantage of the current BOLS approach.

### 3.3 $T_m$ elevation of chemically capped nanosolids

Figure 4a shows the superheating of the chemically capped nanosolids. According to eq (2), superheating of the embedded systems of In/Al ($T_{m,In}/T_{m,Al}$ = 530/932), Pb/Al (600/932), Pb/Zn (600/692), and Ag/Ni (1235/1726)[45] originates from the interfacial bond strengthening. It is understandable that an atom performs differently at a free surface from an atom at the interface. Although the coordination ratio at the interfaces suffers little change ($z_i/z \sim 1$), formation of the interfacial compound or alloy alters the nature of the interatomic bond that should be stronger. Trends show that superheating happens to substances covered by relatively higher $T_m$ substances, or stronger binding systems, as the $T_m$ relates directly to the atomic cohesion energy. Numerical fit leads to a $\alpha$ value of 1.8, indicating that an interfacial bond is 80% stronger than a bond in the core or bulk interior. If we take the bond contraction, 0.90 ~ 0.92, as determined from the As and Bi doped CdTe compound[38] into consideration, it is readily found that the m value is around 5.5 ~7.0. The high m value indicates that bond nature indeed evolves from a compound with m around four to covalent nature. Therefore, the deformed and shortened interfacial bond is much stronger, meaning that electrons at an interface are deeply trapped. Densification of energy, charge, and mass also happens as a result at an interface. Therefore, it is understandable that twins of nanograins[46] and the multilayered structures[47] are stronger and thermally more stable. It is anticipated that a thin insulating layer could form at a junction interface because of the interfacial bond nature alteration and the charge trapping effect. Interestingly, recent theoretical calculations, confirmed by electron microscopy,[48] revealed that homo-junction dislocations in aluminum can have compact or dissociated core interlayers. The calculated minimum stress ($\sigma_P$) required to move an edge dislocation is approximately 20 times higher for compact dislocations than for equivalent dissociated dislocations. As anticipated, this finding provides new insights into the deformation of ultra-fine-grained metals and the twin grain boundaries. Density-functional simulations at temperatures near the $T_m(\infty)$ suggest that the solid-liquid phase-transition temperature at the semiconductor surfaces can be altered via a monolayer coating with a different lattice-matched semiconducting material. A single-monolayer GaAs coating on a Ge(110) surface could raise the Ge melting temperature (1211 K) with an association of a dramatic drop of the diffusion coefficient of the Ge atoms to prevent melting of the bulk Ge layers. In contrast, a single-monolayer coating of Ge on a GaAs(110) surface introduces defects into the bulk and induces a 300 K drop of the $T_m$ of the top layer of GaAs (1540 K). Therefore, superheating is subject to the configuration of the hetero-junction interface and their respective $T_m(\infty)$ as well.

3.4 $T_m$ oscillation

Figure 4b compares the model prediction with the measured $T_m$ change of $Sn^+$, $Al^+$ and $Ga^+$ clusters. It is seen that the $T_m$ curves drop universally at $K_j > 3$ ($Log(K_j) > 0.5$, or $z_i > 3$) disregarding the m values, and then the $T_m$ curves bend up, indicating a higher m value on which the transition point depends, being the same to the m and CN dependence of atomic cohesive energy, as shown in Figure 1. For an isolated atom ($K_j = 0.5$ or $z_i = 0$), the $T_m$ is zero. As the smallest clusters are not in spherical shape and they may subject to structural erratic variation, the equivalent size specified here should be subject to some modification.

The number ratio of surface atoms shows oscillation features with size and the mean $z_i$ values of atoms in magic number clusters are relatively higher because of the compact and closed shell structures.[49] Nevertheless, it is encouraging that the prediction with a m(z) transition from 7 ($z_i = 2$) to 1 ($z_i > 4$) in a form of

$$m(z) = 1 + 12/\{1 + \exp[(z-2)/1.5]\},$$

(3)

matches closely to the measurement of $Ga^+_{17}$, $Ga^+_{39-40}$, $Sn_{10-31}$, and $Sn_{500}$ clusters and Sn nanosolids deposited on $Si_3N_4$ substrate,[50] unfortunately the $Al^+_{49-60}$ clusters. If the $T_m$ rise originates from the $c_i$ deviation without bond nature change, the bond will contract at the lower end of size limit to $c_i = 0.7^7 = 0.082$ that is strictly forbidden (less than 10% of the original bond length!). Therefore, the m value, or bond nature, must change substantially with CN reduction even though the magic number effect is considered. It is hence clear that bond nature of Sn-Sn and Ga-Ga indeed evolves, from metallic-covalent to pure covalent as atomic CN reduces to much lower values ($z_i < 3$), agreeing with that uncovered by Chacko et al.[17] This finding also complies with theoretical findings that the Al-Al bond for lower-coordinated or distorted Al atoms at grain boundaries[51] and at free surfaces[52] become shorter (~5%) and stronger with some more covalent characteristics.[53] However, bond nature evolution (m ~ 2) in $Al^+_{49-60}$ clusters appears not as that significant as occurred in Sn and Ga clusters, as the $T_m$ for $Al^+_{49-63}$ is 300 K lower than the bulk $T_m(\infty)$ of Al (932 K). The abrupt $T_m$ rise (~180 K) for $Al^+_{51-54}$ clusters[14] may partly due to the closed shell structures that are highly stable.[54] Bond nature evolution should also cause conductor-insulator transition such as Pd solid containing $10^{1-2}$ atoms[55] because of the depressed potential well of traps as all the involved atoms are lower-coordinated. The trapping potential well suppression hinders the charge transportation under external electric field.[56] As demonstrated in ref. 27, strong localization of charges in the surface region should take the responsibility for bond nature evolution or alteration, band gap production,[25] and conductivity reduction of small specimens. Results indicate that it is stronger for an IV-A atom to bond with two neighbors than to bond with three or more atoms due to the bond nature evolution, which may explain why a $C_{13}$ cluster prefers a ring or a tadpole structure with each atom two bonds, as theoretically predicted by Ho and coworkers,[18] rather than the densely packed tetrahedron structure. It is evident that the bond nature evolution is the unique property of the III-A and IV-A elements particularly for atoms with larger number of electrons as compared with the m value at the smallest sizes of Al (~ 2), Ga(6 ~ 7), C(m = 2.56), Si(m = 4.88), and Sn (turns from 1 to 6 ~ 7 as the size is reduced).

## IV    Conclusion

In summary, the BOLS correlation premise has enabled the observed $T_m$, $T_C$, and $T_{vap}$ suppression of nanosolids, the $T_m$ elevation of chemically capped nanosolids, and the $T_m$ oscillation over the whole range of solid sizes to be reconciled to the effect of bond-order loss and its consequences on the bond-strength gain or bond-nature evolution. The modified cohesive energy of the lower–coordinated system also determines the geometrical reconstruction, surface lattice or phonon instability, surface energy, or entropy, heat of fusion, and related terms that are used in other theoretical models. From the perspective of equilibration between the thermal energy of phase transition and the cohesive energy of an atom at different sites, the current BOLS premise could favor and incorporate well with existing models to the effect of atomic *CN* imperfection. Consistent understanding further evidences the significance of atomic CN imperfection and the essentiality of the BOLS correlation to the understanding of surface, nanosolids, defects and even amorphous states that involves randomly distributed CN imperfection. Most strikingly, uncovering of the bond nature alteration of the deformed and shortened interfacial bonds should provide deeper insight into the surfaces and interfaces that form the foundation of surface, interface, and nanosolid sciences.

Figure captions

Figure 1 BOLS correlation shows the atomic-CN dependence of the bond length and its effect on the atomic cohesive energy, $E_C = z_i E_i$, that varies with the m value. Scattered data are from Goldschmidt [31] and Feibelman [32]. At high m value, the $E_C$ bend up at a certain $z_i$ values.

Figure 2 Agreement between predictions (lines) and experimental observations (scattered symbols) of the size-and-shape dependence of the $T_m$ suppression of (a) Sn and Al on $Si_3N_4$ substrate,[57,58,59,60] (b) In[61,62] and Pb,[60,63] (c) Au on W,[64] C,[65] and embedded in Silica,[10] (d) Si[66,67] and Ge[68] measured at the beginning and ending of melting, and recrystalizing, (e) Bi[60,61,69,70] and CdS,[71] (f) Ne,[66] Kr,[72] O and Methyl chloride (m-CL).[73]

Figure 3 (a) Magnetic $T_C$ suppression of Co and Ni and (b) evaporation $T_{eva}$ of Ag and PbS nanosolids,[74] indicating consistent trend of the relative change disregarding the actual processes of thermally stimulated phenomena.

Figure 4 (a) Superheating of In[60,75] and Pb[76,77,78,79] embedded in Al matrix.[45] For embedded system, the $z_{ib} c_i^{-m}$ is replaced with a constant α that describes the interfacial bond strength gain. (b) $T_m$ oscillation of those measured from $Ga^+_{13-17}$[15,17], $Ga^+_{39-40}$ [15], $Al^+_{49-60}$ [14], $Sn^+_{10-19}$[21], $Sn^+_{19-31}$ [16], $Sn^+_{500}$ [80] and Sn nanosolid on $Si_3N_4$ substrate. The $T_m$ deviation of $Al^+_{50-60}$ clusters from the predictions indicates (1 < m <3) that bond nature alteration indeed happens to Al but it is less significant compared to Sn and Ga bonds. Numerical fit (solid line) for $Sn^+$ and $Ga^+$ clusters is realized with a function of m(z) = 1+12/{1+exp[(z-2)/1.5]} to let m transit from 7 (at z = 2) to 1(at z > 4).

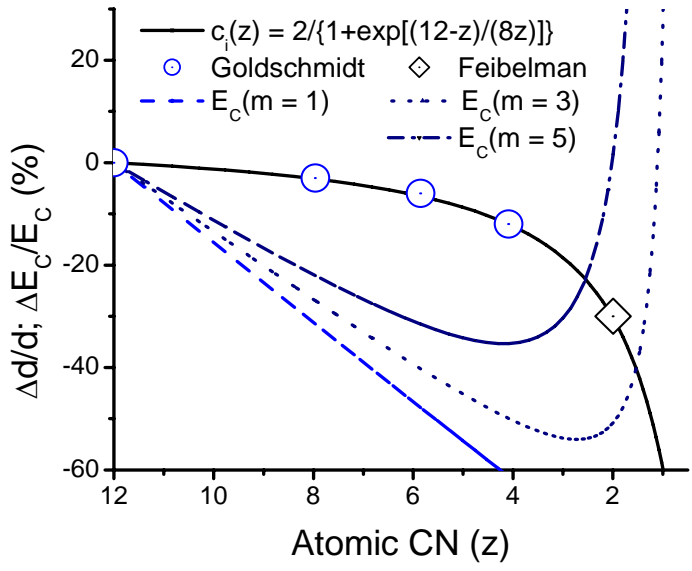

Fig-1 (color on line)

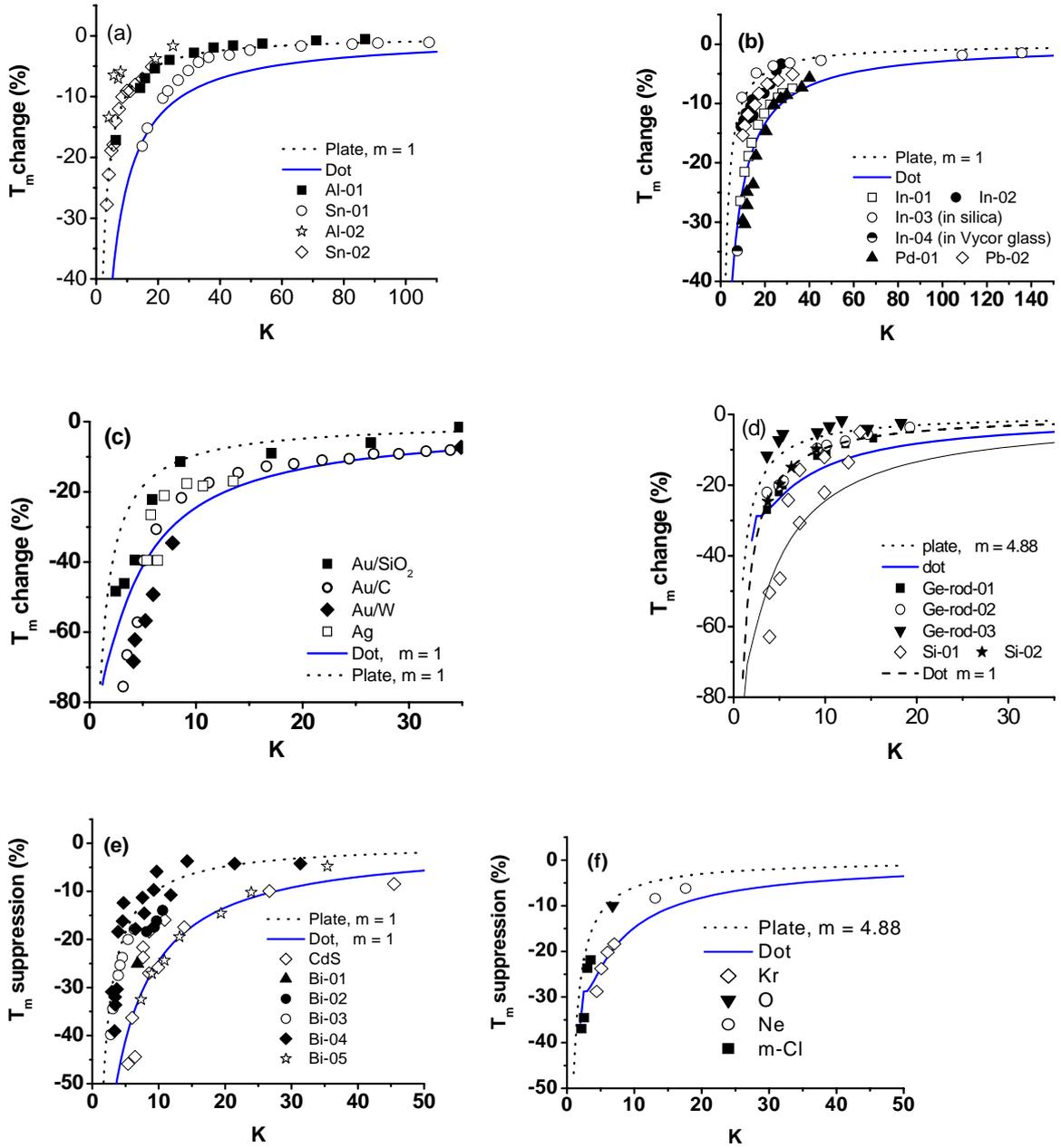

Fig-2

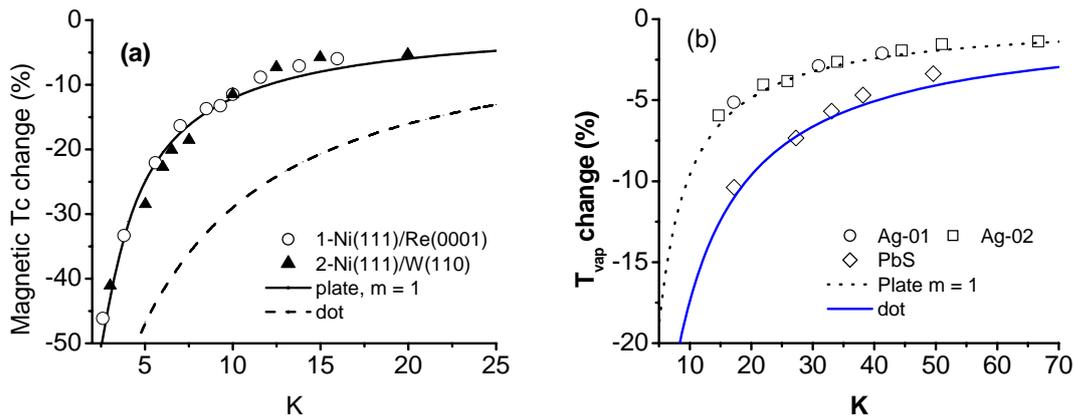

Fig-3

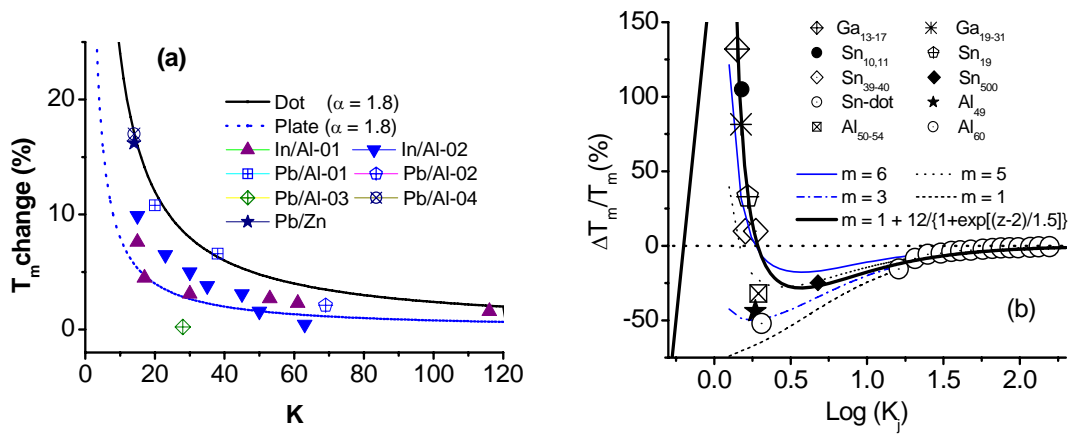

Fig-4

E-mail: ecqsun@ntu.edu.sg; Fax: 65 6792 0415; URL: http://www.ntu.edu.sg/home/ecqsun/